\documentclass[twocolumn,showpacs,amsmath,amssymb]{revtex4}
\usepackage[utf8]{inputenc}
\usepackage{color}
\usepackage{amsmath}
\usepackage{graphicx}
\usepackage{esint}

\makeatletter
\@ifundefined{textcolor}{}
{%
 \definecolor{BLACK}{gray}{0}
 \definecolor{WHITE}{gray}{1}
 \definecolor{RED}{rgb}{1,0,0}
 \definecolor{GREEN}{rgb}{0,1,0}
 \definecolor{BLUE}{rgb}{0,0,1}
 \definecolor{CYAN}{cmyk}{1,0,0,0}
 \definecolor{MAGENTA}{cmyk}{0,1,0,0}
 \definecolor{YELLOW}{cmyk}{0,0,1,0}
}

\makeatother

\begin{document}
\title{Improving estimation of entropy production rate for run-and-tumble
particle systems by high-order thermodynamic uncertainty relation }
\author{Ruicheng Bao}
\author{Zhonghuai Hou}
\thanks{E-mail: hzhlj@ustc.edu.cn}
\affiliation{Department of Chemical Physics \& Hefei National Laboratory for Physical
Sciences at Microscales, iChEM, University of Science and Technology
of China, Hefei, Anhui 230026, China}
\date{\today}
\begin{abstract}
Entropy production plays an important role in the regulation and stability
of active matter systems, and its rate quantifies the nonequilibrium
nature of these systems. However, entropy production is hard to be
experimentally estimated even in some simple active systems like molecular
motors or bacteria, which may be modeled by the run-and-tumble particle
(RTP), a representative model in the study of active matters. Here
we resolve this problem for an asymmetric RTP in one dimension, firstly
constructing a finite time thermodynamic uncertainty relation (TUR)
for an RTP, which works well in the short observation time regime
for entropy production estimation. Nevertheless, when the activity
dominates, i.e., the RTP is far from equilibrium, the lower bound
for entropy production from TUR turns to be trivial. We address this
issue by introducing a recently proposed high-order thermodynamic
uncertainty relation (HTUR), in which the cumulant generating function
of current serves as a key ingredient. To exploit the HTUR, we adopt
a novel method to analytically obtain the cumulant generating function
of the current we study, with no need to explicitly know the time-dependent
probability distribution. The HTUR is demonstrated to be able to estimate
the steady state energy dissipation rate accurately because the cumulant
generating function covers higher-order statistics of the current,
including rare and large fluctuations besides its variance. Compared
to the conventional TUR, the HTUR could give significantly improved
estimation of energy dissipation, which can work well even in the
far-from equilibrium regime. We also provide a strategy based on the
improved bound to estimate the entropy production from moderate amount
of trajectory data for experimental feasibility. 
\end{abstract}
\maketitle

\section{Introduction}

Active matter systems consist of self-propelled particles which can
perform persistent random motion through consuming energy from the
environment and converting it into a nonequilibrium drive \cite{16RMP_active,08PRL_RTP,16PRL_active,19PRX_irreversiblility,18PRE_Hidden}.
In the past two decades, active matters have attracted a surge of
interest in the field of statistical and biological physics, for it
may appropriately model living things like bacteria or flocking birds,
which are far from equilibrium \cite{95PRL_Viscek,95PRL_Tu,98PRE_Tu,14PRE_Szamel,08PRL_RTP,16PRL_active,19PRE_AOUP,19PRE_Szamel,19PRX_irreversiblility,18PRE_Hidden,21PRL_Odd,23PRE_Julicher}.
See \cite{16RMP_active} for a good review. There are three most commonly
concerned active matter models, active Brownian particle (ABP) model
\cite{16PRL_active,20PRL_Gonnella,22hydrodynamic}, active Ornstein-Uhlenbeck
particles (AOUP) model \cite{21JSM_Gonnella,21PRE_Cates} and run-and-tumble
particle (RTP) model \cite{08PRL_RTP,18Fodor}. ABP, AOUP and RTP
are all characterized by active forces with exponential correlations
imposed on them, and all exhibit some nontrivial behaviors compared
to their passive counterparts consequently even at the single particle
level \cite{22PRE_RTP}. For example, an active particle (ABP, AOUP
or RTP) trapped in a confined potential can reach a non-Boltzmann
and non-Gaussian stationary state \cite{19PRE_2dRTP,20PRE_harmonicRTP,21JSM_Gonnella},
and the probability density of an RTP or a AOUP confined in a box
could concentrate near the spatial boundaries \cite{20PRE_BOXRTP,21JSM_Gonnella}.

The entropy production plays a central role in active matter systems,
quantifying the heat dissipation to environment in a steady state,
in other words, quantifying the thermodynamic cost to maintain a nonequilibrium
steady state for some time. However, measuring the entropy production
of these systems directly in experiments is quite challenging because
the temperature changes from dissipation are very small and usually
elusive in the noisy environment \cite{21PRX_ImprovedTUR}. A possible
solution to this issue is the Harada-Sasa relation, which quantitatively
connects the entropy production rate with the violation of the fluctuation-dissipation
relation \cite{05PRL_HSR,16PRL_HSR}. However, it's necessary to measure
whole frequency-spectrum of the focused degree of freedom for the
use of this relation, requiring a lot of statistics. 

Recently, a fundamental inequality called thermodynamic uncertainty
relation (TUR) has been built for general stationary Markov processes,
demonstrating a trade-off relation between precision of an arbitrary
current $j_{\tau}$ (ratio between its squared mean and variance)
and the total entropy production rate $\dot{\Sigma}$ \cite{15PRL_TUR,16PRL_TUR,17PRE_finiteTUR,17PRE_SeifertTUR,19PRE_CRITUR,dechant2018current}:
\begin{equation}
\frac{2k_{B}\langle j_{\tau}\rangle^{2}}{\text{Var}(j_{\tau})}\leq\dot{\Sigma}\tau,
\end{equation}
where $\tau$ is the observation time of the current $j_{\tau}$ (from
now on, we set $k_{B}=1$ for notation brevity, rendering entropy
dimensionless). On top of that, TUR signifies that a lower bound for
the steady state entropy production can be established in terms of
the first and second moment of any currents, which has the potential
to serve as a technique to estimate entropy production only from a
moderate amount of experimentally accessible trajectory data \cite{22PRR_shortTUR,22JCP_CZY,21PRX_ImprovedTUR,20PRL_Short}.
Nonetheless, the estimation from TUR is usually not accurate since
the lower bound is not guaranteed to be tight in general. For instance,
it has been demonstrated that TURs for general biochemical oscillations
are far from tight in several important models like circadian clock
and Brusselator \cite{20PRR_CZY}. For that reason, the accurate estimation
of entropy production from available data in active matter systems
is still an important open problem.

In the present work, we analytically study one of the minimal models
of active matter, the RTP model, providing some useful strategies
to estimate the entropy production rate of a RTP only from trajectories
data obtainable from direct experimental observations. To begin, we
have built a finite-time TUR for this model, showing that the lower
bound of entropy production given by the TUR serves as a good estimator
in the short observation time limit. Nevertheless, very high temporal
resolution is needed to keep the estimation robust when the activity
is large. To address this, we build a tighter lower bound of entropy
production as a new estimator by incorporating the effect of large
and rare fluctuations. This new estimator is based on the recently
proposed high-order thermodynamic uncertainty relation (HTUR) \cite{dechant2020fluctuation,HighTUR,21PRR_CTR},
which is robust when the RTP is arbitrarily far from equilibrium and
the observation time is not short. The key quantity of the HTUR is
the cumulant generating function (CGF) of the interested current.
To exploit the HTUR, we provide a novel approach to analytically calculate
the CGF directly from the Fokker-Planck equations of the system. In
consequence, the HTUR bound can be directly evaluated through our
exact expression of the CGF. We also propose an experimentally practical
strategy to get better estimation of entropy production than conventional
TUR since the CGF may not be possible to obtain in experiment. Our
work provides some insight on the HTUR, which may find further applications
in other active matter systems.

The rest of the paper is organized as follows. In section II, we introduce
the asymmetric RTP model. In section III, a finite-time TUR is constructed
analytically for this model, and the transport efficiency which show
the TUR's performance of estimating entropy production is evaluated
under different observation times and activities. In section IV, the
HTUR for the RTP is derived, which is utilized to significantly improve
the estimation of entropy production, followed by Section V with conclusions
and outlooks.

\section{Model}

\begin{figure}
\begin{centering}
\includegraphics[width=1\columnwidth]{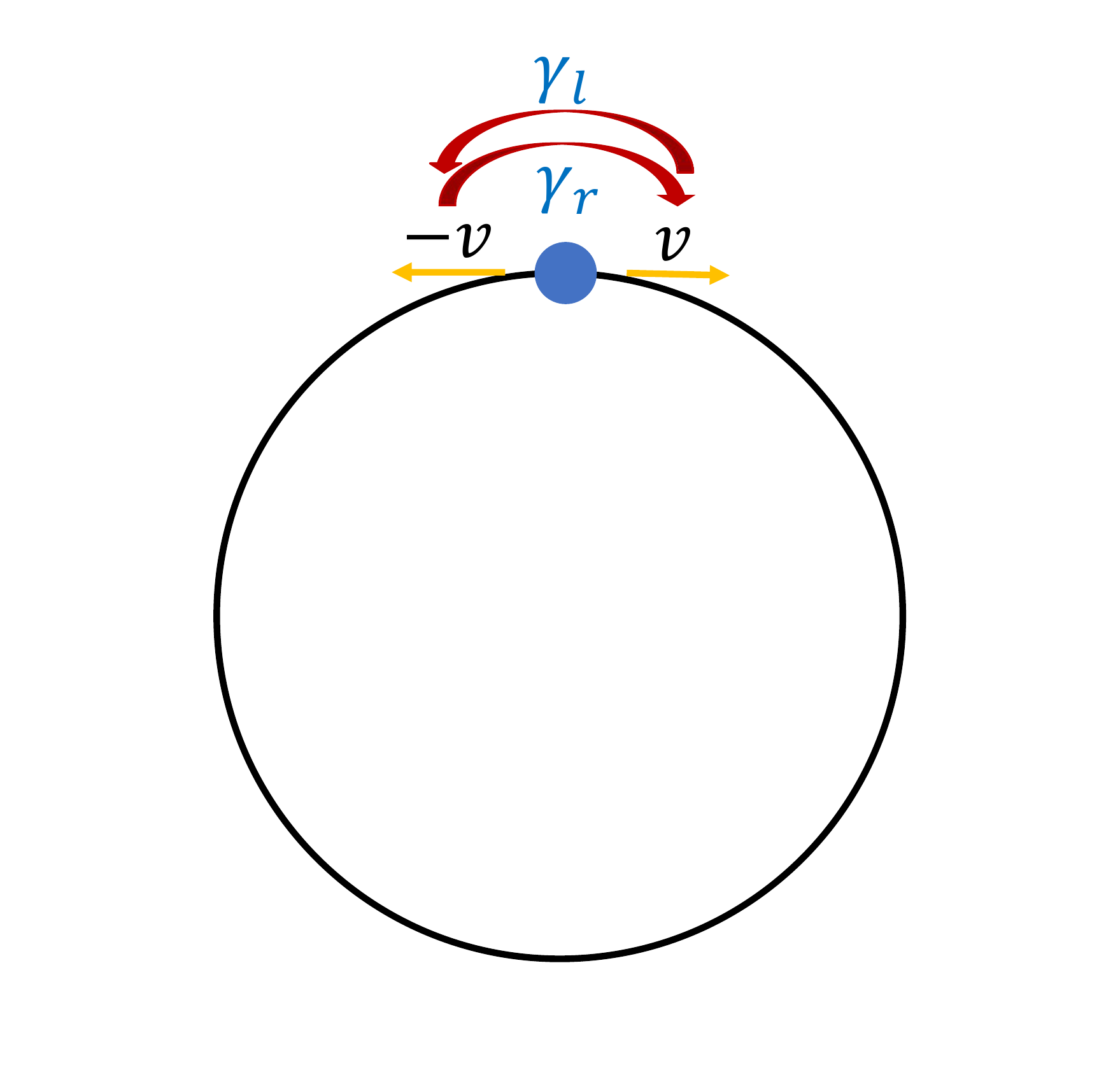}
\par\end{centering}
\caption{An illustration of a run-and-tumble particle moving along a one-dimensional
ring whose circumference is $L$.}

\label{fig1}
\end{figure}

Throughout this work, we consider an asymmetric one-dimensional RTP
model with diffusion. In this model, the position of a single RTP
is described by the Langevin equation (the mobility $\mu$ is set
to be $1$) 
\begin{equation}
\frac{dx}{dt}=v\sigma(t)+\sqrt{2D}\xi(t),\label{Langevin}
\end{equation}
where $D=T\mu=T$ is the diffusion constant, $v$ is a constant drift
velocity, $\xi(t)$ is the Gaussian white noise with zero mean $\langle\xi(t)\rangle=0$
and delta-function correlation $\langle\xi(t)\xi(s)\rangle=\delta(t-s)$
and $\sigma(t)=\pm1$ refers to a dichotomous telegraphic noise that
switches from the run-state to the tumble-state at rate $\gamma_{r}$
and at rate $\ensuremath{\gamma_{l}}$ conversely. Note that $\sigma(t)$
is a colored noise whose stationary auto-correlation function is given
by {[}see Appendix A for proof{]} 
\begin{equation}
\langle\sigma(t)\sigma(s)\rangle=\frac{4\gamma_{r}\gamma_{l}}{(\gamma_{r}+\gamma_{l})^{2}}e^{-\left(\gamma_{r}+\gamma_{l}\right)\lvert t-s\rvert}+\left(\frac{\gamma_{r}-\gamma_{l}}{\gamma_{r}+\gamma_{l}}\right)^{2}.\label{correlation}
\end{equation}
The corresponding Fokker-Planck equation of Eq.(\ref{Langevin}) reads
\begin{align}
\frac{\partial p_{r}(x,t)}{\partial t} & =-\frac{\partial j_{r}(x,t)}{\partial x}-\gamma_{r}p_{r}(x,t)+\gamma_{l}p_{l}(x,t)\label{FPEr}\\
\frac{\partial p_{l}(x,t)}{\partial t} & =-\frac{\partial j_{l}(x,t)}{\partial x}+\gamma_{r}p_{r}(x,t)-\gamma_{l}p_{l}(x,t),\label{FPEl}
\end{align}
where the probability currents $j_{r,l}(x,t)$ are defined as 
\begin{align*}
j_{r}(x,t) & =\left[v-D\partial_{x}\right]p_{r}(x,t)\\
j_{l}(x,t) & =\left[-v-D\partial_{x}\right]p_{l}(x,t),
\end{align*}
and $p_{r}(x,t)$ ($p_{l}(x,t)$) denotes the probability of finding
a particle with velocity $v$ ($-v$) at position $x$ and time $t$.
Without loss of generality, we assume $\gamma_{l}>\gamma_{r}\geq0$
so that the RTP would move along the same direction as the drift velocity
$v$ on average. To assure ergodicity, the RTP is set to be confined
in a one-dimensional ring whose circumference is $L$, i.e., $x\in[0,L)$,
so that the stationary state distribution is the uniform distribution
$p^{st}(x)=\text{\ensuremath{\lim_{t\rightarrow\infty}p(x,t)}}=1/L$.
What's more, the stationary distribution of the particle being in
run-state and tumble-state are $p_{r}^{st}(x)=\gamma_{l}/(\gamma_{l}+\gamma_{r})L$
and $p_{l}^{st}(x)=\gamma_{r}/(\gamma_{l}+\gamma_{r})L$ respectively.
However, we claim that for natural boundary condition, the scheme
to estimate dissipation in this paper still works in the large time
limit. This can be understood by noticing that the natural boundary
condition is effectively the periodic boundary condition with $L\rightarrow\infty$
for one-dimensional systems, and the entropy production rate in the
large time limit and its estimator in this work is irrelevant to the
system size $L$ (see Appendix B). A schematic illustration of our
model is given in Figure \ref{fig1}. Very recently, Ro et. al. have
experimentally studied the entropy production of a four-state run-and-tumble
particle jumping along a ring, which is analogous to our model \cite{22PRL_EPmeasure}.

We would like to explain why we study the RTP with asymmetric transition
rates between the run-state and the tumble-state. Molecular motors
and \textit{Escherichia coli} in nature usually exhibit directed motion.
To better model this directed motion, one should consider asymmetric
transition rates instead of symmetric transition rates in which case
$\gamma_{r}=\gamma_{l}=\gamma$. In the symmetric case, the RTP won't
display directed movement because the mean displacement $\langle x(\tau)\rangle_{x_{0}}-x_{0}$
always vanishes in the large time limit whatever the initial position
$x_{0}$ is, i.e., $\langle x(\tau)\rangle_{x_{0}}=x_{0}$, just like
the unbiased random walk. Here, $\langle x(\tau)\rangle_{x_{0}}$
is the mean position of the particle at time $t=\tau$ with the initial
position at $t=0$ being given by $x_{0}$.

\section{Finite-time thermodynamic uncertainty relation for a run-and-tumble
particle}

\begin{figure}
\begin{centering}
\includegraphics[width=1\columnwidth]{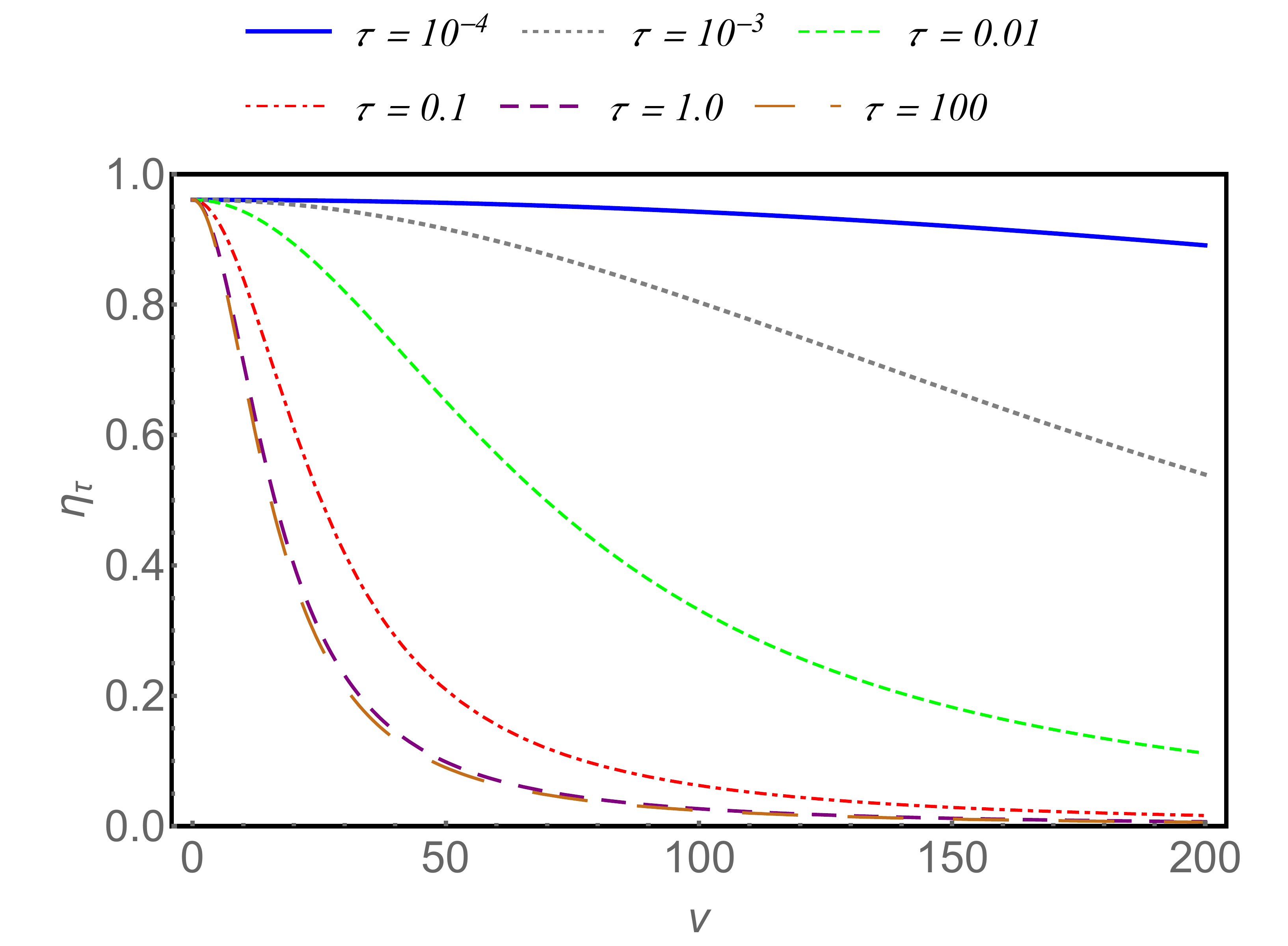}
\par\end{centering}
\caption{The transport efficiency $\eta_{\tau}$ from the TUR bound versus
the activity $v$ with different observation time $\tau$. The model
parameters are chosen as $\gamma_{l}=10,\ \gamma_{r}=0.1,\ D=1.0.$}

\label{fig2}
\end{figure}

In the beginning, we show the validity of the conventional TUR and
its limitation on the estimation of energy dissipation in our model.
Below, we construct a TUR and use it to estimate the steady state
entropy production rate. To study the TUR, we consider a fluctuating
generalized current defined as 
\begin{equation}
j_{\tau}=\int_{x(0)}^{x(\tau)}w(x)\circ\dot{x}(t)\text{d}t,\label{current}
\end{equation}
where $w(x)$ is a differentiable weight function and $\circ$ denotes
the Stratonovich product. We choose $w(x)=1$ so that the resulting
current is a most physically relevant quantity, i.e., the displacement
during the finite observation time $\tau$. Note that the choice of
mean particle displacement as the focused current is aimed to assure
experimentally easy accessibility. In the stationary state, the mean
value of this current $x_{\tau}\equiv x(\tau)-x(0)=\int_{0}^{\tau}\dot{x}(t)dt$
can be readily obtained as 
\begin{equation}
\langle x_{\tau}\rangle=\int_{0}^{\tau}\langle\dot{x}\rangle dt=v\int_{0}^{\tau}\langle\sigma(t)\rangle dt=\frac{\gamma_{l}-\gamma_{r}}{\gamma_{l}+\gamma_{r}}v\tau,\label{meanvalue}
\end{equation}
and its variance in the steady state can also be computed as (see
Appendix B for derivations) 
\begin{align}
\text{Var}(x_{\tau})= & D_{\text{eff}}\tau-\frac{8\gamma_{l}\gamma_{r}v^{2}}{(\gamma_{l}+\gamma_{r})^{4}}\left[1-e^{-(\gamma_{l}+\gamma_{r})\tau}\right],\nonumber \\
D_{\text{eff}}\equiv & \frac{8\gamma_{l}\gamma_{r}v^{2}}{(\gamma_{l}+\gamma_{r})^{3}}+2D.\label{effDiff}
\end{align}
Note that the variance can be calculated via the noise correlation
(\ref{correlation}), or through the moment equations method introduced
in the Appendix B. Since the displacement is nothing but the accumulation
of instantaneous velocity during $\tau$, $x(\tau)$ in our spatially
periodic model is the same as its counterpart in the model with natural
boundary condition. Equipped with expressions of the mean and variance,
an estimator of the steady state mean entropy production during an
observation interval $\tau$ may be obtained according to the conventional
TUR: 
\begin{align}
\Sigma_{\text{TUR}}^{\tau}= & \frac{2\langle x_{\tau}\rangle^{2}}{\text{Var}(x_{\tau})}\label{TUR}\\
= & \frac{2\left(\gamma_{l}^{2}-\gamma_{r}^{2}\right)^{2}v^{2}\tau}{D_{\text{eff}}(\gamma_{l}+\gamma_{r})^{4}-\frac{8\gamma_{l}\gamma_{r}v^{2}[1-e^{-(\gamma_{l}+\gamma_{r})\tau}]}{\tau}}\leq\langle\Sigma_{\tau}\rangle,\nonumber 
\end{align}
which provides a lower bound of the steady state entropy production.
Below, the steady state mean entropy production is calculated exactly
by stochastic thermodynamics \cite{seifert2012stochastic}, so that
the Eq.(\ref{TUR}) can be easily verified, i.e., 
\begin{align}
\langle\Sigma_{\tau}\rangle= & \tau\left(\int_{0}^{L}\frac{j_{r}^{st}(x)^{2}}{Dp_{r}^{st}(x)}dx+\int_{0}^{L}\frac{j_{l}^{st}(x)^{2}}{Dp_{l}^{st}(x)}dx\right)\nonumber \\
+ & \tau\int_{0}^{L}dx\left[\gamma_{r}p_{r}^{st}(x)-\gamma_{l}p_{l}^{st}(x)\right]\ln\frac{\gamma_{r}p_{r}^{st}(x)}{\gamma_{l}p_{l}^{st}(x)}\nonumber \\
= & \frac{v^{2}\tau}{D}\left(\frac{\gamma_{l}}{\gamma_{r}+\gamma_{l}}+\frac{\gamma_{r}}{\gamma_{r}+\gamma_{l}}\right)=\frac{v^{2}\tau}{D},\label{EPR}
\end{align}
where in the second line, the relation $\gamma_{r}p_{r}^{st}(x)-\gamma_{l}p_{l}^{st}(x)=0$
has been used. The expression for the steady state entropy production
is the same as a diffusive particle with constant drift $v$, because
the RTP can be regarded as a diffusive particle with drift $v$ ceaselessly
changing direction instantaneously, and these instant changes of direction
won't produce entropy \cite{cocconi2020entropy}. Thus, it's obvious
from Eq.(\ref{EPR}) that the TUR (\ref{TUR}) is validated. Then
a transport efficiency 
\begin{align}
\eta_{\tau} & \equiv\frac{\Sigma_{\text{TUR}}^{\tau}}{\langle\Sigma\rangle}\nonumber \\
= & \frac{\left(\gamma_{l}^{2}-\gamma_{r}^{2}\right)^{2}}{\frac{D_{\text{eff}}}{2D}\left(\frac{\gamma_{r}+\gamma_{l}}{\gamma_{r}-\gamma_{l}}\right)^{2}-\frac{8\gamma_{l}\gamma_{r}v^{2}[1-e^{-(\gamma_{l}+\gamma_{r})\tau}]}{2\left(\gamma_{l}^{2}-\gamma_{r}^{2}\right)^{2}D\tau}}\leq1
\end{align}
can be defined to evaluate the efficiency of estimation \cite{dechant2018current,hwang2018energetic}.
To illustrate the estimating effect of TUR, we plot the transport
efficiency from TUR with different observation time $\tau$ and different
activity $v$ in Figure \ref{fig2}. 

From the above expression and plot we can draw some conclusions. First
of all, if the drift velocity $v$ is very large, i.e., the RTP is
far from equilibrium, meanwhile the observation time $\tau$ is large
for experimental convenience, then the TUR bound will become very
loose and therefore cannot work well for the entropy production estimation,
whatever other system details are. It has been reported that the TUR
can be tight in the linear-response regime, but generally it will
be loose when the system is far from equilibrium \cite{15PRL_TUR,16PRL_TUR,HighTUR}.
Physically, this is because the presence of excess fluctuations (quantified
by high-order cumulants) far from equilibrium \cite{21PRR_CTR}. In
the linear-response regime these fluctuations are negligible so that
the TUR can be tight. Secondly, in the short observation time limit
($\tau\rightarrow0$) the TUR estimator would work remarkably well
compared to the large observation time cases: 
\begin{align}
\lim_{\tau\rightarrow\infty}\eta_{\tau}= & \frac{2D}{D_{\text{eff}}}\left(\frac{\gamma_{r}-\gamma_{l}}{\gamma_{r}+\gamma_{l}}\right)^{2}\nonumber \\
\leq & \lim_{\tau\rightarrow0}\eta_{\tau}=\left(\frac{\gamma_{r}-\gamma_{l}}{\gamma_{r}+\gamma_{l}}\right)^{2}\leq1,
\end{align}
whose estimating effect would be robust even in the far from equilibrium
region ($v\gg1$), since it is irrelevant to the drift velocity $v$.
The inequality saturates when $\gamma_{r}=0$, i.e., the particle
won't tumble, but always move forward with a constant mean velocity.
When $\gamma_{r}=\gamma_{l},$ TUR becomes trivial and cannot gives
any prediction, because the mean displacement vanishes, making the
TUR estimator vanish as well. The robustness under activity is in
fact the advantage of the recently found short-time TUR \cite{20PRL_Short,20PRE_shortTUR}.
The short-time TUR is tight even in the far from equilibrium regime,
for the reason that the short observation time kills excess fluctuations
(interested readers can refer to supplemental material of \cite{20PRL_Short}
for details). However, experimental scientists may prefer large measurement
time TUR as an estimating method, due to the limitation of time-resolution
of common tools. Here we would like to address this issue utilizing
the HTUR.

\section{High-order thermodynamic uncertainty relation and its application
to estimate entropy production}

Recently, Dechant and Sasa \cite{dechant2020fluctuation} constructed
a HTUR from their fluctuation-response inequality in both Langevin
systems and discrete-state Markov systems , which reads 
\begin{align}
\langle x_{\tau}\rangle^{2}\sup_{h}F(h)\leq & \langle\Sigma_{\tau}\rangle,\label{HTUR}\\
F(h):= & \frac{h^{2}}{K_{x_{\tau}}(h)-h\langle x_{\tau}\rangle}\nonumber 
\end{align}
where $K_{x_{\tau}}(h)\equiv\ln\langle e^{hx_{\tau}}\rangle$ is the
CGF of the current $x_{\tau}$ (this inequality works for generalized
current $j_{\tau},$ in this work we only focus on $x_{\tau}$). The
conventional TUR can be readily recovered from Eq.(\ref{HTUR}) by
taking the $h\rightarrow0$ limit. It has been demonstrated that in
a general jump-diffusion model \cite{dechant2020fluctuation}, Eq.
(\ref{HTUR}) still works (see Appendix C for details). This type
of process can be described by an equation 
\begin{equation}
\dot{x}(t)=a_{k(t)}[x(t)]+\sqrt{2D_{k(t)}}\xi(t),\label{jpmodel}
\end{equation}
where the drift term $a_{k(t)}[x(t)]$ and diffusion coefficient $D_{k(t)}$
can jump between multiple discrete states $k=1,...,N$. The jumping
dynamics is described by a Markov jump process with transition rates
$W_{ij}$ from state $j$ to state $i$. This model covers our RTP
as a specific case, thus the HTUR can be applied to our model.

To enhance the estimation of entropy production on account of HTUR,
we calculate the CGF of the current $x_{\tau}$ below. We define some
quantities for after use: 
\begin{align}
\langle e^{hx(\tau)}\rangle_{r} & \equiv\int e^{hx}p_{r}(x,\tau)dx\\
\langle e^{hx(\tau)}\rangle_{l} & \equiv\int e^{hx}p_{l}(x,\tau)dx,
\end{align}
so that 
\begin{equation}
\langle e^{hx(\tau)}\rangle=\langle e^{hx(\tau)}\rangle_{r}+\langle e^{hx(\tau)}\rangle_{l}.
\end{equation}
It can be readily demonstrated that {[}see Appendix D for proof{]}
\begin{equation}
\langle e^{hx_{\tau}}\rangle=\langle e^{hx(\tau)}\rangle_{x(0)=0},\label{Equality}
\end{equation}
which is in accordance with physical intuition since the initial position
$x(0)$ is extracted from the uniform steady state distribution (any
two points in our stationary system are identical). Then one can directly
write down the evolution equations for $\langle e^{hx(\tau)}\rangle_{r,l}$
by multiplying $e^{hx}$ on both sides of the Fokker-Planck equations
(\ref{FPEr})-(\ref{FPEl}) and integrating over the whole range of
$x$, i.e., 
\begin{align}
\frac{d\langle e^{hx(\tau)}\rangle_{r}}{d\tau} & =\left(Dh^{2}+vh-\gamma_{r}\right)\langle e^{hx(\tau)}\rangle_{r}+\gamma_{l}\langle e^{hx(\tau)}\rangle_{l}\\
\frac{d\langle e^{hx(\tau)}\rangle_{l}}{d\tau} & =\left(Dh^{2}-vh-\gamma_{l}\right)\langle e^{hx(\tau)}\rangle_{l}+\gamma_{r}\langle e^{hx(\tau)}\rangle_{r}.
\end{align}
Above equations can be rewritten as a compact vector form 
\begin{equation}
\frac{d\vec{\phi}(\tau)}{d\tau}=\mathcal{L}\vec{\phi}(\tau),\label{linear}
\end{equation}
where 
\begin{align*}
\vec{\phi}(\tau) & \equiv(\langle e^{hx(\tau)}\rangle_{r},\langle e^{hx(\tau)}\rangle_{l})^{\text{T}}
\end{align*}
and 
\[
\mathcal{L}=\left(\begin{array}{cc}
Dh^{2}+vh-\gamma_{r} & \gamma_{l}\\
\gamma_{r} & Dh^{2}-vh-\gamma_{l}
\end{array}\right).
\]
Eq.(\ref{linear}) is linear, thus its solution can be formally written
as 
\begin{equation}
\vec{\phi}(\tau)=e^{\mathcal{L}\tau}\vec{\phi}(0),\label{formal}
\end{equation}
with the initial condition being ($x(0)=0$) 
\[
\vec{\phi}(0)=\left(\frac{\gamma_{l}}{\gamma_{l}+\gamma_{r}},\frac{\gamma_{r}}{\gamma_{l}+\gamma_{r}}\right)^{\text{T}}.
\]
Then the closed form of $K_{x_{\tau}}(h)$ can be obtained by 
\begin{equation}
K_{x_{\tau}}(h)=\ln\left(\langle e^{hx(\tau)}\rangle_{r}+\langle e^{hx(\tau)}\rangle_{l}\right),
\end{equation}
which may be too lengthy to be placed here. We include the detailed
form of it in the Appendix D for completeness. However, in the short
observation time limit $\tau\rightarrow0$, the expression of the
CGF is brief (high-order terms may be dropped due to their negligible
effects in the maximization problem): 
\begin{equation}
K_{x_{\tau}}(h)=\langle x_{\tau}\rangle h+Dh^{2}\tau+\mathcal{O}(\tau^{2}),
\end{equation}
resulting in the same lower bound as the short-time TUR: 
\begin{equation}
\Sigma_{\text{HTUR}}^{\tau\rightarrow0}=\left(\frac{\gamma_{r}-\gamma_{l}}{\gamma_{r}+\gamma_{l}}\right)^{2}\frac{v^{2}}{D}\tau=\frac{\langle x_{\tau}\rangle^{2}}{D\tau}.\label{short}
\end{equation}
Additionally, the leading contribution $K_{\text{lead}}(h)$ of $K_{x_{\tau}}(h)$
in the large $\tau$ limit could be identified as 
\begin{equation}
\frac{2Dh^{2}+\sqrt{\left(\gamma_{l}+\gamma_{r}\right)^{2}-4hv(\gamma_{l}-\gamma_{r})+4h^{2}v^{2}}-\gamma_{l}-\gamma_{r}}{2}\tau,
\end{equation}
then maximize 
\begin{equation}
\frac{h^{2}}{\text{\ensuremath{K_{\text{lead}}}}(h)-h\langle x_{\tau}\rangle}
\end{equation}
over the whole range of $h$ remarkably gives rise to $1/D\tau$,
when $h\rightarrow\infty$. This still leads to the tight bound (\ref{short})
as in the small $\tau$ limit. In other cases with $0<\tau<\infty$
the optimization problem $\sup_{h}F(h)$ from HTUR might not be solved
generally when parameters are not fixed. In spite of this, we discover
that for any $\tau$ the $h\rightarrow\infty$ limit for $F(h)$ can
be obtained as (see Appendix D for details)
\begin{equation}
\lim_{h\rightarrow\infty}F(h)=\frac{1}{D\tau},
\end{equation}
which means that the lower bound $\Sigma_{\text{HTUR}}^{\tau}$ given
by HTUR cannot be smaller than $\frac{1}{D\tau}$ , because the lower
bound is given by the maximal value of $F(h)$. That is, we have 
\begin{equation}
\langle\Sigma_{\tau}\rangle\geq\Sigma_{\text{HTUR}}^{\tau}=\langle x_{\tau}\rangle^{2}\sup_{h}F(h)\geq\frac{\langle x_{\tau}\rangle^{2}}{D\tau}.
\end{equation}
After numerically exploring a large amount of values over the parameter
space $(\gamma_{l},\gamma_{r},v,D,\tau)$, we claim that $F(h)=\frac{h^{2}}{K_{x_{\tau}}(h)-h\langle x_{\tau}\rangle}$
is an increasing function of $h$ when $h>0$, and when $h<0$ the
function $F(h)<F(-h)$ (see Appendix E for numerical evidence). Based
on the above findings, we conjecture that a new lower bound for entropy
production from HTUR for any observation time $\tau$ is given by
\begin{equation}
\Sigma_{\text{HTUR}}^{\tau}=\frac{\langle x_{\tau}\rangle^{2}}{D\tau}.\label{mainresult}
\end{equation}
which is our main result.

\begin{figure}
\begin{centering}
\includegraphics[width=1\columnwidth]{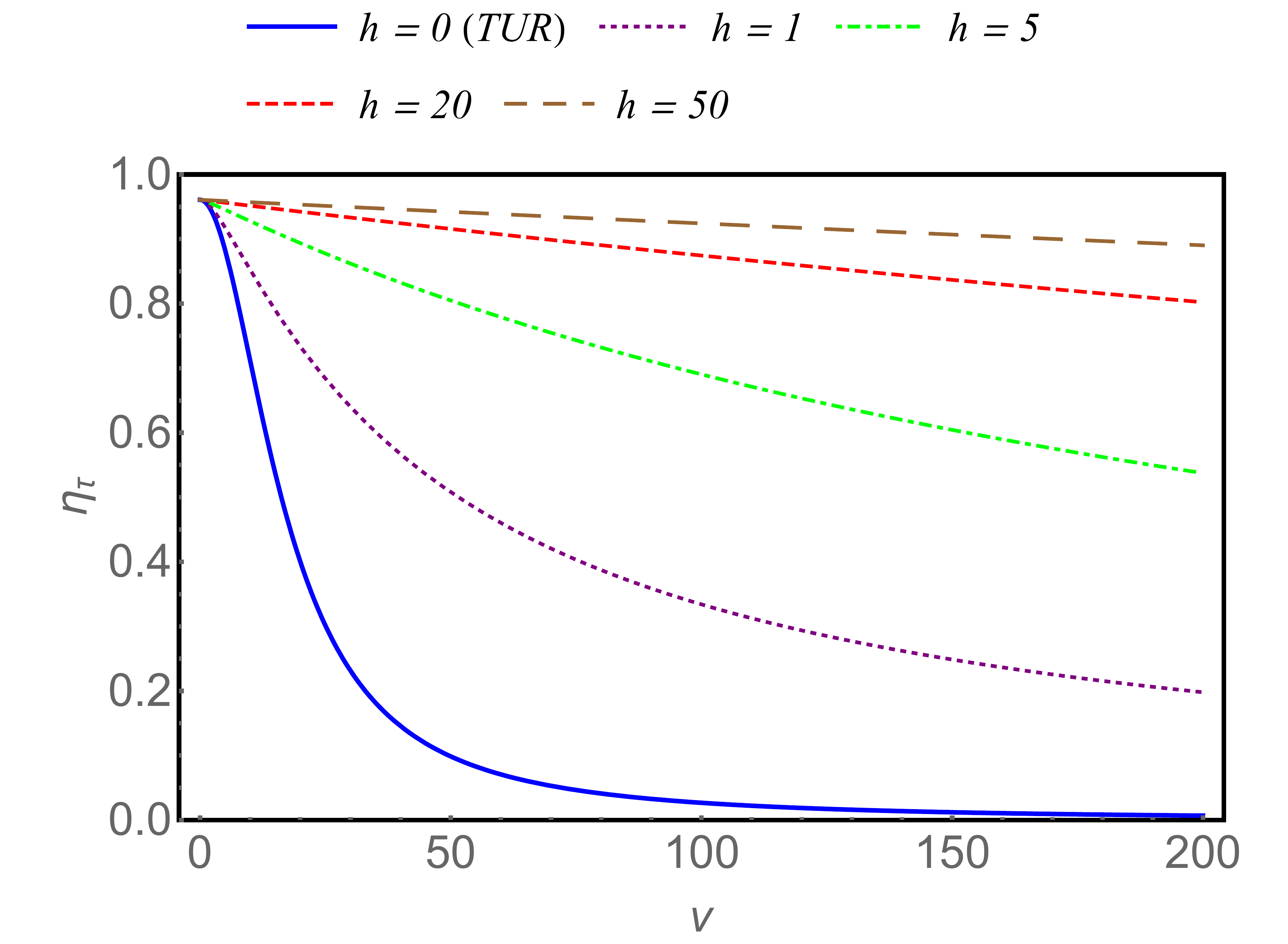}
\par\end{centering}
\caption{The transport efficiency $\eta_{\tau}=\Sigma_{h=a}^{\tau}/\langle\Sigma_{\tau}\rangle$
from the TUR bound and from the improved bound versus the activity
$v$ with different $h$. The model parameters are chosen as $\gamma_{l}=10,\ \gamma_{r}=0.1,\ D=1.0,\ \tau=1.0$ }

\label{fig3}
\end{figure}

Some remarks on this result can be made. Firstly, the estimation of
entropy production rate from Eq.(\ref{mainresult}) would not be affected
by the variation in observation time $\tau$, and would be robust
even in the far from equilibrium region, in stark contrast to the
conventional TUR. Secondly, when the diffusion constant $D$ (or the
friction coefficient) is known, the energy dissipation during $\tau$
can be estimated experimentally only by readily measuring the mean
displacement $\langle x_{\tau}\rangle$ during that time interval.
Therefore, the HTUR estimator may find potential application in many
active matter systems, since the amount of trajectory data needed
for is pretty small compared to other methods. However, the system
details are usually unknown, one would prefer to measure the dissipation
only through the trajectory information, in which case the CGF should
be measured to obtain our tighter bound. Notwithstanding the CGF which
contains the information of infinite higher-order cumulants may not
be experimentally feasible, we show in Figure \ref{fig3} that when
the factor $h$ is fixed, the left-hand-side of Eq.(\ref{HTUR}) can
still serve as a pretty good estimator for entropy production, i.e.,
\begin{equation}
\Sigma_{h=a}^{\tau}\equiv\frac{a^{2}\langle x_{\tau}\rangle^{2}}{\ln\langle e^{ax_{\tau}}\rangle-a\langle x_{\tau}\rangle}\leq\langle\Sigma_{\tau}\rangle.
\end{equation}
And when $h$ is fixed, the resulting estimator could be experimentally
obtained from the time series data of trajectories, without prior
knowledge of the model details. Strikingly, as shown in Figure \ref{fig3},
even the estimator 
\begin{equation}
\Sigma_{h=1}^{\tau}=\frac{\langle x_{\tau}\rangle^{2}}{\ln\langle e^{x_{\tau}}\rangle-\langle x_{\tau}\rangle}
\end{equation}
in $h=1$ case would greatly improve the estimation of entropy production
compared to the conventional TUR, behaving much better in the far
from equilibrium regime. When the chosen values of $h$ are increasing,
the resulting estimators become better and better, and asymptotically
converge to the best one $\Sigma_{\text{HTUR}}^{\tau}$. 

\section{Discussion}

In this paper, we explore the stochastic thermodynamics of an asymmetric
run-and-tumble particle, which may model behaviors of molecular motors
or chemotaxis motions of some active bacteria. We firstly explore
the finite-time TUR in our system, revealing that the short observation
time strategy is beneficial for the estimation of entropy production.
Most importantly, resorting to the HTUR, we have shown that an improved
estimation of energy dissipation only from experimentally feasible
trajectory data can be realized. The HTUR estimating strategy is robust
when the RTP is arbitrarily far from equilibrium, and its effect won't
be affected by the observation time $\tau$, forming a sharp contrast
to the conventional TUR. Based on the HTUR, we further propose an
experimentally viable estimating strategy for entropy production rate,
and check its effect through the analytical expression of CGF, showing
that the strategy still significantly outperform the conventional
TUR strategy. We would like to emphasize the advantage of our estimating
strategy based on TUR or HTUR. The chosen current observable can be
measured on the very coarse-grained level, so that only a moderate
amount of trajectory data is required. To apply our strategy, there
is even no need to track the whole trajectory of the position $x$.
For each experiment, measurements of the number of cycles the particle
goes through during the observation time, the initial position and
the final position of the particle at time $\tau$ are enough for
estimation, with detecting the current state of the particle (run-state
or tumble-state) being unnecessary. To apply our estimating method,
the requirement for the spatial and temporal resolution of experimental
equipment is relatively low. Therefore, we reveal the potential strength
of HTUR in the estimation of entropy production in active matter systems.

There are still some limitations of our work. The asymmetry of the
hopping rate between run-state and tumble-state is necessary for our
estimators both from conventional TUR and HTUR, due to the choice
of displacement as the current to use. If one chooses the entropy
production itself as a current, the TUR and HTUR bound can be saturated
even in the symmetric case $\gamma_{l}=\gamma_{r}$\cite{16PRL_TUR,20PRE_shortTUR,20PRL_Short},
in which our bound cannot be applied. Nevertheless, our aim is to
estimate the entropy production of RTP, once the entropy production
itself has been known, it's no need to do any estimation any more.
Therefore, whether there is a good estimator in the symmetric case
$\gamma_{l}=\gamma_{r}$ still remains to be an open problem, which
we leave for future work. What's more, when the RTP is trapped in
a confined potential (see Apendix B for an example), the mean displacement
vanishes in the stationary state, in which case our estimator cannot
take effect either. Besides, the generalization of our method to two-dimensional
RTP is nontrivial and deserves for further study. 
\begin{acknowledgments}
This work is supported by MOST(2018YFA0208702) and NSFC (21833007). 
\end{acknowledgments}

\appendix
\onecolumngrid

\section*{Appendix}


\section{A simple derivation of Eq.(\ref{correlation})}

From the definition of stationary average, we directly write down
\begin{align}
\langle\sigma(t)\sigma(s)\rangle= & \sum_{\sigma_{1},\sigma_{2}=\pm1}\sigma_{1}\sigma_{2}p(\sigma_{2},t|\sigma_{1},s)p^{st}(\sigma_{1})\\
= & \sum_{\sigma_{1}}\langle\sigma(t)\rangle_{\sigma_{1},s}\sigma_{1}p^{st}(\sigma_{1}),\label{colornoise}
\end{align}
where $\langle\sigma(t)\rangle_{\sigma_{1},s}$ is the conditional
average with the initial condition being $\sigma(s)=\sigma_{1}$.
From the Fokker-Planck equation $p(\sigma,t|\sigma_{1},s)$ obeys
and the initial condition $p(\sigma,s|\sigma_{1},s)=\delta_{\sigma,\sigma_{2}}$,
one can solve that 
\begin{align}
p(\sigma & =1,t|\sigma_{1},s)=\frac{\gamma_{l}}{\gamma_{l}+\gamma_{r}}\nonumber \\
+ & e^{-(\gamma_{l}+\gamma_{r})(t-s)}\left(\frac{\gamma_{r}}{\gamma_{l}+\gamma_{r}}\delta_{1,\sigma_{1}}-\frac{\gamma_{l}}{\gamma_{l}+\gamma_{r}}\delta_{-1,\sigma_{1}}\right),\\
p(\sigma & =-1,t|\sigma_{1},s)=\frac{\gamma_{r}}{\gamma_{l}+\gamma_{r}}\nonumber \\
- & e^{-(\gamma_{l}+\gamma_{r})(t-s)}\left(\frac{\gamma_{r}}{\gamma_{l}+\gamma_{r}}\delta_{1,\sigma_{1}}-\frac{\gamma_{l}}{\gamma_{l}+\gamma_{r}}\delta_{-1,\sigma_{1}}\right).
\end{align}
Then the conditional average can be computed as 
\begin{align}
\langle\sigma(t)\rangle_{\sigma_{1},s}= & \sum_{\sigma=\pm1}\sigma(t)p(\sigma,t|\sigma_{1},s)\nonumber \\
= & \frac{\gamma_{l}-\gamma_{r}}{\gamma_{l}+\gamma_{r}}+e^{-(\gamma_{l}+\gamma_{r})(t-s)}\left(\sigma_{1}-\frac{\gamma_{l}-\gamma_{r}}{\gamma_{l}+\gamma_{r}}\right).\label{conditionave}
\end{align}
Plugging Eq.(\ref{conditionave}) into Eq.(\ref{colornoise}) and
using 
\begin{align*}
p^{st}(\sigma & =1)=\frac{\gamma_{l}}{\gamma_{l}+\gamma_{r}}\\
p^{st}(\sigma & =-1)=\frac{\gamma_{r}}{\gamma_{l}+\gamma_{r}}
\end{align*}
one obtains that 
\begin{equation}
\langle\sigma(t)\sigma(s)\rangle=\frac{4\gamma_{r}\gamma_{l}}{(\gamma_{r}+\gamma_{l})^{2}}e^{-\left(\gamma_{r}+\gamma_{l}\right)\lvert t-s\rvert}+\left(\frac{\gamma_{r}-\gamma_{l}}{\gamma_{r}+\gamma_{l}}\right)^{2},
\end{equation}
which is just the Eq.(\ref{correlation}) in the main text.

\section{Moment equations of the run-and-tumble particle in one dimension}

In this appendix, we introduce the moment equations method and give
some applications.

\subsubsection*{Calculation of variance of the current $x_{\tau}$}

In this subsection, we use the moment equations method to obtain the
variance for the one-dimensional RTP, which would be useful in the
main text. Firstly, let's define some quantities which would be useful.
The $n^{th}$ right moment, left moment and moment of the position
at time $t$ are respectively given by 
\begin{align}
\langle x^{n}\rangle_{r} & \equiv\int x(t)^{n}p_{r}(x,t)dx,\nonumber \\
\langle x^{n}\rangle_{l} & \equiv\int x(t)^{n}p_{l}(x,t)dx,\nonumber \\
\langle x^{n}\rangle & =\langle x^{n}\rangle_{r}+\langle x^{n}\rangle_{l}=\int x(t)^{n}p(x,t)dx.
\end{align}
Multiplying $x^{n}$ on both sides of Eq.(\ref{FPEr}) and (\ref{FPEl})
one obtains the evolution equations for the $n^{th}$ right moment
and left moment as 
\begin{align}
\frac{d\langle x^{n}\rangle_{r}}{dt} & =\theta_{n,1}nv\langle x^{n-1}\rangle_{r}+\theta_{n,2}Dn(n-1)\langle x^{n-2}\rangle_{r}-\gamma_{r}\langle x^{n}\rangle_{r}+\gamma_{l}\langle x^{n}\rangle_{l}\\
\frac{d\langle x^{n}\rangle_{l}}{dt} & =-\theta_{n,1}nv\langle x^{n-1}\rangle_{l}+\theta_{n,2}Dn(n-1)\langle x^{n-2}\rangle_{r}+\gamma_{r}\langle x^{n}\rangle_{r}-\gamma_{l}\langle x^{n}\rangle_{l},
\end{align}
where $\theta_{n,c}$ equals $1$ for $n\geq c$ and $0$ for $n<c$.
When $n=0,$ the above equations reduce to 
\begin{align}
\frac{d\langle x^{0}\rangle_{r}}{dt} & =-\gamma_{r}\langle x^{0}\rangle_{r}+\gamma_{l}\langle x^{0}\rangle_{l}\label{n0}\\
\frac{d\langle x^{0}\rangle_{l}}{dt} & =\gamma_{r}\langle x^{0}\rangle_{r}-\gamma_{l}\langle x^{0}\rangle_{l}.\label{n0b}
\end{align}
In the large time limit that we are interested in, combining Eq.(\ref{n0})-(\ref{n0b})
with the conservation of probability $\langle x^{0}\rangle_{r}+\langle x^{0}\rangle_{l}=1$
gives rise to $\langle x^{0}\rangle_{r}=\gamma_{l}/(\gamma_{l}+\gamma_{r})$
and $\langle x^{0}\rangle_{l}=\gamma_{r}/(\gamma_{l}+\gamma_{r})$.
When $n=1$, the first-order moment equations are 
\begin{align}
\frac{d\langle x\rangle_{r}}{dt} & =v\frac{\gamma_{l}}{\gamma_{l}+\gamma_{r}}-\gamma_{r}\langle x\rangle_{r}+\gamma_{l}\langle x\rangle_{l}\\
\frac{d\langle x\rangle_{l}}{dt} & =-v\frac{\gamma_{r}}{\gamma_{l}+\gamma_{r}}+\gamma_{r}\langle x\rangle_{r}-\gamma_{l}\langle x\rangle_{l},
\end{align}
which leads to 
\begin{equation}
\frac{d\langle x\rangle}{dt}=\frac{d\langle x\rangle_{r}}{dt}+\frac{d\langle x\rangle_{l}}{dt}=v\frac{\gamma_{l}-\gamma_{r}}{\gamma_{l}+\gamma_{r}}.\label{n1}
\end{equation}
From the initial condition $\langle x(0)\rangle=\langle x\rangle^{st}=L/2$,
the first moment at time $\tau$ is yielded: 
\begin{equation}
\langle x(\tau)\rangle=\langle x\rangle_{r}+\langle x\rangle_{l}=v\frac{\gamma_{l}-\gamma_{r}}{\gamma_{l}+\gamma_{r}}\tau+\frac{L}{2},\label{1moment}
\end{equation}
Combined Eq.(\ref{1moment}) with the first-order equations, the first
left and right moment at time $\tau$ can also be expressed as 
\begin{align}
\langle x\rangle_{r} & =\frac{\gamma_{l}L}{2(\gamma_{l}+\gamma_{r})}+\frac{\gamma_{l}\left[\left(\gamma_{l}^{2}-\gamma_{r}^{2}\right)\tau+2\gamma_{r}\right]v}{(\gamma_{l}+\gamma_{r})^{3}}-\frac{2\gamma_{l}\gamma_{r}ve^{-(\gamma_{l}+\gamma_{r})\tau}}{(\gamma_{l}+\gamma_{r})^{3}}\label{nr1}\\
\langle x\rangle_{l} & =\frac{\gamma_{r}L}{2(\gamma_{l}+\gamma_{r})}+\frac{\gamma_{r}\left[\left(\gamma_{l}^{2}-\gamma_{r}^{2}\right)\tau-2\gamma_{l}\right]v}{(\gamma_{l}+\gamma_{r})^{3}}+\frac{2\gamma_{l}\gamma_{r}ve^{-(\gamma_{l}+\gamma_{r})\tau}}{(\gamma_{l}+\gamma_{r})^{3}},\label{nl1}
\end{align}
having taken the initial conditions (steady state) 
\begin{equation}
\langle x_{0}\rangle_{r}=\frac{L}{2}\frac{\gamma_{l}}{\gamma_{l}+\gamma_{r}},\ \langle x_{0}\rangle_{l}=\frac{L}{2}\frac{\gamma_{r}}{\gamma_{l}+\gamma_{r}}
\end{equation}
into account. Then from the second-order moment equations ($n=2$),
one can figure out the variance of $x(\tau)$, which reads 
\begin{align}
\frac{d\langle x^{2}\rangle_{r}}{dt} & =2v\langle x\rangle_{r}+2D\langle x^{0}\rangle_{r}-\gamma_{r}\langle x^{2}\rangle_{r}+\gamma_{l}\langle x^{2}\rangle_{l}\\
\frac{d\langle x^{2}\rangle_{l}}{dt} & =-2v\langle x\rangle_{l}+2D\langle x^{0}\rangle_{l}+\gamma_{r}\langle x^{2}\rangle_{r}-\gamma_{l}\langle x^{2}\rangle_{l}.
\end{align}
Thus the second moment of $x(\tau)$ arises from the equation 
\begin{equation}
\frac{d\langle x^{2}\rangle}{dt}=2v\left(\langle x\rangle_{r}-\langle x\rangle_{l}\right)+2D,
\end{equation}
whose solution is 
\begin{equation}
\langle x(\tau)^{2}\rangle=\left(\langle x(\tau)\rangle^{2}-\frac{L^{2}}{4}\right)+D_{\text{eff}}\tau-\frac{8\gamma_{l}\gamma_{r}v^{2}}{(\gamma_{l}+\gamma_{r})^{4}}\left[1-e^{-(\gamma_{l}+\gamma_{r})\tau}\right]+\frac{L^{2}}{3}
\end{equation}
with the initial condition being $\langle x(0)^{2}\rangle=L^{2}/3$
and Eq.(\ref{nr1})-(\ref{nl1}) being used. Here, the effective diffusion
coefficient has been defined in the Eq.(\ref{effDiff}) of main text
as 
\[
D_{\text{eff}}=\frac{8\gamma_{l}\gamma_{r}v^{2}}{(\gamma_{l}+\gamma_{r})^{3}}+2D.
\]
So the variance of $x(\tau)$ is simply 
\begin{align}
\text{Var}(x(\tau))= & \langle x(\tau)^{2}\rangle-\langle x(\tau)\rangle^{2}\nonumber \\
= & D_{\text{eff}}\tau-\frac{8\gamma_{l}\gamma_{r}v^{2}}{(\gamma_{l}+\gamma_{r})^{4}}\left[1-e^{-(\gamma_{l}+\gamma_{r})\tau}\right]+\frac{L^{2}}{12}
\end{align}
In the stationary state, the variance of the current $x_{\tau}$ with
observation time $\tau$ is connected to the variance of $x(\tau)$
as

\begin{align}
\text{Var}(x_{\tau})= & \langle\left(x(\tau)-x(0)\right)^{2}\rangle-\left(\langle x(\tau)\rangle-\langle x(0)\rangle\right)^{2}\nonumber \\
= & \langle x(\tau)^{2}\rangle+\langle x(0)^{2}\rangle-2\langle x(\tau)x(0)\rangle-\langle x(\tau)\rangle^{2}-\langle x(0)\rangle^{2}+2\langle x(\tau)\rangle\langle x(0)\rangle\nonumber \\
= & \text{Var}(x(\tau))+L^{2}/12-2\text{Cov}[x(\tau),x(0)],
\end{align}
where $\text{Var}(x(0))=L^{2}/3-(L/2)^{2}=L^{2}/12$ have been used.
Now let's compute the quantity $\text{Cov}[x(\tau),x(0)]=\langle x(\tau)x(0)\rangle-\langle x(\tau)\rangle\langle x(0)\rangle.$
Note that 
\begin{align}
\langle x(\tau)x(0)\rangle= & \int dx\int_{0}^{L}dx_{0}x(\tau)x_{0}p(x,\tau|x_{0},0)p^{st}(x_{0})dxdx_{0}\\
= & \frac{1}{L}\int_{0}^{L}\langle x(\tau)\rangle_{x_{0}}x_{0}dx_{0},
\end{align}
with $\langle x(\tau)\rangle_{x_{0}}\equiv\int x(\tau)p(x,\tau|x_{0},0)dx.$
According to Eq.(\ref{n1}), 
\begin{equation}
\langle x(\tau)\rangle_{x_{0}}=v\frac{\gamma_{l}-\gamma_{r}}{\gamma_{l}+\gamma_{r}}\tau+x_{0},
\end{equation}
so that 
\begin{align}
\langle x(\tau)x(0)\rangle-\langle x(\tau)\rangle\langle x(0)\rangle= & \frac{1}{L}\int_{0}^{L}\langle x(\tau)\rangle_{x_{0}}x_{0}dx_{0}-\frac{v\tau L}{2}\frac{\gamma_{l}-\gamma_{r}}{\gamma_{l}+\gamma_{r}}-\frac{L^{2}}{4}\nonumber \\
= & \frac{L^{2}}{12}.
\end{align}
Therefore, the variance of the current $x_{\tau}$ is finally obtained
as 
\begin{align}
\text{Var}(x_{\tau})= & \text{Var}(x(\tau))+L^{2}/12-2\text{Cov}[x(\tau),x(0)]\nonumber \\
= & \text{Var}(x(\tau))-\frac{L^{2}}{12}\nonumber \\
= & D_{\text{eff}}\tau-\frac{8\gamma_{l}\gamma_{r}v^{2}}{(\gamma_{l}+\gamma_{r})^{4}}\left[1-e^{-(\gamma_{l}+\gamma_{r})\tau}\right],
\end{align}
which is just the Eq.(\ref{effDiff}) of the main text. Note that
the expression of moments of $x_{\tau}$ is irrelevant to the system
size $L$, thus it can also be applied to the natural boundary condition
case. It should be mentioned that this result can also be directly
derived from the two-time correlation function of the velocity $\dot{x}(t)=v\sigma(t)+\sqrt{2D}\xi(t)$,
using the celebrated Green-Kubo (G-K) formula. We sketch the derivation
using G-K formula below. Since 
\begin{equation}
\langle x_{\tau}^{2}\rangle=\langle\int_{0}^{\tau}\int_{0}^{\tau}\dot{x}(t)\dot{x}(s)dsdt\rangle=\int_{0}^{\tau}\int_{0}^{\tau}\langle\dot{x}(t)\dot{x}(s)\rangle dsdt,
\end{equation}
what we need to compute is the two-time correlation function 
\begin{align}
C(t-s)\equiv & \langle\dot{x}(t)\dot{x}(s)\rangle=v^{2}\langle\sigma(t)\sigma(s)\rangle+2D\langle\xi(t)\xi(s)\rangle\nonumber \\
= & v^{2}\left[\frac{4\gamma_{r}\gamma_{l}}{(\gamma_{r}+\gamma_{l})^{2}}e^{-\left(\gamma_{r}+\gamma_{l}\right)\lvert t-s\rvert}+\left(\frac{\gamma_{r}-\gamma_{l}}{\gamma_{r}+\gamma_{l}}\right)^{2}\right]+2D\delta(t-s).
\end{align}
Then according to the G-K formula, 
\begin{align}
\langle x_{\tau}^{2}\rangle= & \int_{0}^{\tau}\int_{0}^{\tau}C(t-s)dsdt=2\int_{0}^{\tau}C(t)(\tau-t)dt\nonumber \\
= & D_{\text{eff}}\tau-\frac{8\gamma_{l}\gamma_{r}v^{2}}{(\gamma_{l}+\gamma_{r})^{4}}\left[1-e^{-(\gamma_{l}+\gamma_{r})\tau}\right]+\langle x_{\tau}\rangle^{2},
\end{align}
so that 
\[
\text{Var}(x_{\tau})=D_{\text{eff}}\tau-\frac{8\gamma_{l}\gamma_{r}v^{2}}{(\gamma_{l}+\gamma_{r})^{4}}\left[1-e^{-(\gamma_{l}+\gamma_{r})\tau}\right].
\]
Note that the integration of the delta function $\int_{0}^{\tau}\delta(t)dt=1/2$
here because the Stratonovich convention is taken.

\subsubsection*{Energy dissipation rate of a RTP in a harmonic potential well}

Here, we discuss another application of the moment equations method,
calculating the entropy production rate of a RTP confined in a harmonic
potential $U(x)=\frac{1}{2}kx^{2}$. The result is a slight generalization
of what was obtained in the Reference \cite{21JSM_RTP}, where the
entropy production rate of a symmetric RTP in a harmonic potential
has been calculated through the field-theoretical method. The corresponding
Langevin equation and Fokker-Planck equations are 
\begin{equation}
\dot{x}(t)=-kx+v\sigma(t)+\sqrt{2D}\xi(t)
\end{equation}
and 
\begin{align}
\frac{\partial p_{r}(x,t)}{\partial t}= & \partial_{x}\left[kx-v_{0}+D\partial_{x}\right]p_{r}(x,t)-\gamma_{r}p_{r}(x,t)+\gamma_{l}p_{l}(x,t)\\
\frac{\partial p_{l}(x,t)}{\partial t}= & \partial_{x}\left[kx+v_{0}+D\partial_{x}\right]p_{l}(x,t)+\gamma_{r}p_{r}(x,t)-\gamma_{l}p_{l}(x,t).
\end{align}
In this case, the RTP will finally converge to a nonequilibrium stationary
state (NESS). The moment equations at this stationary state are obtained
as 
\begin{align}
nk\langle x^{n}\rangle_{r} & =\theta_{n,1}nv\langle x^{n-1}\rangle_{r}+\theta_{n,2}Dn(n-1)\langle x^{n-2}\rangle_{r}-\gamma_{r}\langle x^{n}\rangle_{r}+\gamma_{l}\langle x^{n}\rangle_{l}\\
nk\langle x^{n}\rangle_{l} & =-\theta_{n,1}nv\langle x^{n-1}\rangle_{l}+\theta_{n,2}Dn(n-1)\langle x^{n-2}\rangle_{r}+\gamma_{r}\langle x^{n}\rangle_{r}-\gamma_{l}\langle x^{n}\rangle_{l},
\end{align}
where $\langle\cdot\rangle_{r,l}=\int dx(\cdot)p_{r,l}^{st}(x)$ is
the stationary state average. Note that the energy dissipation rate
at the NESS is only contributed by the switching between run-state
and tumble-state, which can be regarded as a potential switching process
\cite{16PRL_HSR} between the left potential $V_{l}(x)\equiv U(x)+vx$
and the right potential $V_{r}(x)\equiv U(x)-vx$. Consequently, the
energy dissipation rate can be readily computed by 
\begin{align}
W= & \int dx\gamma_{r}p_{r}^{st}(x)\Delta V+\int dx\gamma_{l}p_{l}^{st}(x)(-\Delta V)\nonumber \\
= & \int dx\left[\gamma_{r}p_{r}^{st}(x)-\gamma_{l}p_{l}^{st}(x)\right]\left[V_{l}(x)-V_{r}(x)\right]\nonumber \\
= & 2v\int dx\left[\gamma_{r}p_{r}^{st}(x)-\gamma_{l}p_{l}^{st}(x)\right]x\nonumber \\
= & 2v\left[\gamma_{r}\langle x\rangle_{r}-\gamma_{l}\langle x\rangle_{l}\right],
\end{align}
with $\Delta V\equiv V_{l}(x)-V_{r}(x)=2vx.$ This is because in the
presence of a confined potential, the contribution from the drift
vanishes in the stationary state (effective equilibrium). From the
conservation of probability $\langle x^{0}\rangle_{r}+\langle x^{0}\rangle_{l}=1$
we still have $\langle x^{0}\rangle_{r}=\gamma_{l}/(\gamma_{l}+\gamma_{r})$
and $\langle x^{0}\rangle_{l}=\gamma_{r}/(\gamma_{l}+\gamma_{r})$,
then taking $n=1$ in the above stationary state moment equations
brings about 
\begin{align}
k\langle x\rangle_{r} & =\frac{v\gamma_{l}}{\gamma_{l}+\gamma_{r}}-\gamma_{r}\langle x\rangle_{r}+\gamma_{l}\langle x\rangle_{l}\\
k\langle x\rangle_{l} & =\frac{-v\gamma_{r}}{\gamma_{l}+\gamma_{r}}+\gamma_{r}\langle x\rangle_{r}-\gamma_{l}\langle x\rangle_{l}.
\end{align}
These two equations directly lead to 
\begin{align}
\langle x\rangle & =\langle x\rangle_{r}+\langle x\rangle_{l}=\frac{\gamma_{l}-\gamma_{r}}{\gamma_{l}+\gamma_{r}}\frac{v}{k},\nonumber \\
\langle x\rangle_{r} & =\frac{\gamma_{l}}{\gamma_{l}+\gamma_{r}+k}\left(\langle x\rangle+\frac{v}{\gamma_{l}+\gamma_{r}}\right)\nonumber \\
\langle x\rangle_{l} & =\frac{\gamma_{r}}{\gamma_{l}+\gamma_{r}+k}\left(\langle x\rangle-\frac{v}{\gamma_{l}+\gamma_{r}}\right).
\end{align}
As a result, the steady state energy dissipation rate is 
\begin{equation}
W=2v\left[\gamma_{r}\langle x\rangle_{r}-\gamma_{l}\langle x\rangle_{l}\right]=\frac{4v^{2}\gamma_{r}\gamma_{l}}{(\gamma_{l}+\gamma_{r}+k)(\gamma_{l}+\gamma_{r})},
\end{equation}
And the entropy production rate is ($D=T$) 
\begin{equation}
\dot{\Sigma}=\frac{W}{D}=\frac{4v^{2}\gamma_{r}\gamma_{l}}{D(\gamma_{l}+\gamma_{r}+k)(\gamma_{l}+\gamma_{r})},
\end{equation}
reducing to the main result in reference \cite{21JSM_RTP} 
\begin{equation}
\dot{\Sigma}_{\text{sym}}=\frac{v^{2}\alpha}{D(k+\alpha)}
\end{equation}
when $\gamma_{l}=\gamma_{r}=\frac{\alpha}{2}$. In the $k\rightarrow0$
limit, 
\begin{equation}
\lim_{k\rightarrow0}\dot{\Sigma}=\frac{4v^{2}\gamma_{r}\gamma_{l}}{D(\gamma_{l}+\gamma_{r})^{2}},\label{switchingcost}
\end{equation}
which seems to deviate from the real entropy production rate $\frac{v^{2}}{D}$
when there is no confined potential (i.e., when $k$ rigorously equals
to zero). Actually, there is a part of entropy production rate which
is contributed by the nonzero steady state mean velocity ($\bar{v}=\frac{\gamma_{l}-\gamma_{r}}{\gamma_{l}+\gamma_{r}}v$):
\begin{align}
\dot{\Sigma}_{\text{v}}= & \left(p_{r}^{st}F_{r}\cdot\bar{v}+p_{l}^{st}F_{l}\cdot\bar{v}\right)\frac{1}{D}\nonumber \\
= & \frac{\gamma_{l}-\gamma_{r}}{\gamma_{l}+\gamma_{r}}\frac{v}{D}\left(\frac{\gamma_{l}v}{\gamma_{l}+\gamma_{r}}+\frac{\gamma_{r}(-v)}{\gamma_{l}+\gamma_{r}}\right)\nonumber \\
= & \left(\frac{\gamma_{l}-\gamma_{r}}{\gamma_{l}+\gamma_{r}}\right)^{2}\frac{v^{2}}{D},
\end{align}
where $F_{r}\equiv v$ and $F_{l}=-v$ are two constant forces applied
to the RTP with opposite directions. Only when no confined potential
exists, will this part of contribution emerges. When $k$ is not exactly
equal zero, the particle will still be confined in a harmonic potential
so that $\bar{v}=0$ and the contribution $\dot{\Sigma}_{\text{v}}$
vanishes. Adding this contribution to Eq.(\ref{switchingcost}), the
real entropy production rate without confined potential is recovered:
\begin{equation}
\dot{\Sigma}=\frac{v^{2}}{D}\left[\frac{4\gamma_{r}\gamma_{l}}{(\gamma_{l}+\gamma_{r})^{2}}+\left(\frac{\gamma_{l}-\gamma_{r}}{\gamma_{l}+\gamma_{r}}\right)^{2}\right]=\frac{v^{2}}{D}.
\end{equation}
In contrast, no matter how small the value of $k$ is (no matter how
soft the confined potential is), the mean velocity of the RTP in the
stationary state vanishes once there still has a confined potential.
From the above analysis, we can identify the term 
\begin{equation}
\dot{\Sigma}_{\text{sw}}\equiv\frac{4v^{2}\gamma_{r}\gamma_{l}}{D(\gamma_{l}+\gamma_{r})^{2}}
\end{equation}
as the part of entropy production rate originating from state-switching,
which may not experimentally estimated by trajectory data using TUR
or HTUR. That is, though the entropy production rate can be calculated
exactly, it may be difficult to measure it experimentally without
knowing the model details. Therefore, it's still an open problem to
find a experimentally feasible strategy to estimate the entropy production
of a RTP in a confined potential.

\section{High-order TUR in the jump-diffusion model}

In this appendix, we derive the inequality (\ref{HTUR}) for the jump
diffusion model (\ref{jpmodel}) for completeness, following references
\cite{dechant2020fluctuation,21PRR_CTR}. Note that throughout this
appendix, we are only focused on steady states. Firstly, one needs
to introduce a family of dynamics denoted by a parameter $\theta$
\begin{equation}
\dot{x}(t)=a_{k(t)}^{\theta}[x(t)]+\sqrt{2D_{k(t)}}\xi(t),
\end{equation}
with 
\begin{equation}
a_{k(t)}^{\theta}[x(t)]=\theta v^{st}(x)+D_{k(t)}\partial_{x}\ln p^{st}(x),
\end{equation}
The $\theta\in[-1,1]$ has been named as the continuous time-reversal
parameter, affecting the mean current as 
\begin{equation}
\langle J_{\tau}\rangle^{\theta}=\theta\langle J_{\tau}\rangle,
\end{equation}
where $J_{\tau}$ is the generalized current defined in Eq.(\ref{current})
of main text, and $\langle\cdot\rangle^{\theta}=\int dx(\cdot)p^{\theta}(x)$.
Furthermore, the stationary state distribution $p^{st}(x)$ always
keeps unchanged when the value of $\theta$ changes. Considering two
(path) probability densities $p^{\theta_{1}}(x)$ and $p^{\theta_{2}}(x)$
from different dynamics (here $x$ may denote a fluctuating trajectory
$\{x(t)\}_{t\in[0,\tau]}$ from $0$ to $\tau$ or simply a state
variable), one has 
\begin{equation}
K_{J_{\tau}}^{\theta_{1}}(h)=\ln\left(\int dxe^{hJ_{\tau}(x)}p^{\theta_{1}}(x)\right)=\ln\left(\int dxe^{hJ_{\tau}(x)}\frac{p^{\theta_{1}}(x)}{p^{\theta_{2}}(x)}p^{\theta_{2}}(x)\right),
\end{equation}
then from the concavity of logarithm, Jensen inequality tells that
\begin{align}
K_{J_{\tau}}^{\theta_{1}}(h)\geq & \int dx\ln\left(e^{hJ_{\tau}(x)}\frac{p^{\theta_{1}}(x)}{p^{\theta_{2}}(x)}\right)p^{\theta_{2}}(x)\nonumber \\
= & h\langle J_{\tau}\rangle^{\theta_{2}}-D_{\text{KL}}(p^{\theta_{2}}||p^{\theta_{1}}),\label{Kullback}
\end{align}
with the Kullback-Leibler (KL) divergence being defined as 
\begin{equation}
D_{\text{KL}}(p^{\theta_{2}}||p^{\theta_{1}})=\int dxp^{\theta_{2}}(x)\ln\frac{p^{\theta_{2}}(x)}{p^{\theta_{1}}(x)}.
\end{equation}
Since inequality (\ref{Kullback}) holds for any real value of $h$,
it can be rewritten as a lower bound for KL-divergence, which reads
\begin{equation}
D_{\text{KL}}(p^{\theta_{2}}||p^{\theta_{1}})\geq\sup_{h}\left[h\langle J_{\tau}\rangle^{\theta_{2}}-K_{J_{\tau}}^{\theta_{1}}(h)\right].\label{LBKL}
\end{equation}
For the jump-diffusion dynamics, the KL-divergence between the distributions
of two dynamics can be decomposed as 
\begin{equation}
D_{\text{KL}}(p^{\theta_{2}}||p^{\theta_{1}})=D_{\text{KL}}^{\text{diff}}(p^{\theta_{2}}||p^{\theta_{1}})+D_{\text{KL}}^{\text{jump}}(p^{\theta_{2}}||p^{\theta_{1}})+D_{\text{KL}}^{\text{ini}}(p_{0}^{\theta_{2}}||p_{0}^{\theta_{1}}),
\end{equation}
where the first term is the contribution from diffusion part, the
second term from jump part, and the last term from the difference
in two initial distributions. Because we are considering the steady
state, which isn't affected by $\theta$, the last term vanishes.
It has been shown that for path probability densities, 
\begin{align}
D_{\text{KL}}^{\text{diff}}(p^{\theta_{2}}||p^{\theta_{1}}) & =\frac{\left(\theta_{1}-\theta_{2}\right)^{2}}{4}\langle\Sigma_{\tau}^{\text{diff}}\rangle,\\
D_{\text{KL}}^{\text{jump}}(p^{\theta_{2}}||p^{\theta_{1}}) & \leq\frac{\left(\theta_{1}-\theta_{2}\right)^{2}}{4}\langle\Sigma_{\tau}^{\text{jump}}\rangle.
\end{align}
As a consequence, one has 
\begin{align}
D_{\text{KL}}(p^{\theta_{2}}||p^{\theta_{1}})\leq & \frac{\left(\theta_{1}-\theta_{2}\right)^{2}}{4}\left(\langle\Sigma_{\tau}^{\text{diff}}\rangle+\langle\Sigma_{\tau}^{\text{jump}}\rangle\right)\nonumber \\
= & \frac{\left(\theta_{1}-\theta_{2}\right)^{2}}{4}\langle\Sigma_{\tau}\rangle.\label{UBKL}
\end{align}
Combining Eq.(\ref{Kullback}) and (\ref{UBKL}) gives rise to 
\begin{equation}
K_{J_{\tau}}^{\theta_{1}}(h)\geq h\theta_{2}\langle J_{\tau}\rangle-\frac{\left(\theta_{1}-\theta_{2}\right)^{2}}{4}\langle\Sigma_{\tau}\rangle,
\end{equation}
then maximize the right hand side with respect to $\theta_{2}$ resulting
in a quadratic bound under any $\theta_{1}$ 
\begin{equation}
K_{J_{\tau}}^{\theta_{1}}(h)\geq h\theta_{1}\langle J_{\tau}\rangle+\frac{h^{2}\langle J_{\tau}\rangle^{2}}{\langle\Sigma_{\tau}\rangle}.
\end{equation}
Rearranging this, set $\theta=1$ and maximize over the whole range
of $h$, one obtains the HTUR (\ref{HTUR}) 
\begin{equation}
\langle\Sigma_{\tau}\rangle\geq\langle J_{\tau}\rangle^{2}\sup_{h}\frac{h^{2}}{K_{J_{\tau}}(h)-h\langle J_{\tau}\rangle}.
\end{equation}

\section{Calculation of $K_{x_{\tau}}(h)$ and $\lim_{h\rightarrow\infty}F(h)$}

In this appendix, we analytically calculate the cumulant generating
function $K_{x_{\tau}}(h)=\ln\langle e^{hx_{\tau}}\rangle$ of the
current $x_{\tau}$. Firstly, let's prove the equality (\ref{Equality})
in the main text. Denoting $x(0)=x_{0},$ the left hand side of it
is 
\begin{align}
\langle e^{hx_{\tau}}\rangle= & \langle e^{h[x(\tau)-x_{0}]}\rangle=\int dx\int_{0}^{L}dx_{0}e^{h[x(\tau)-x_{0}]}p(x,\text{\ensuremath{\tau}}|x_{0},0)p(x_{0})\\
= & \int dxe^{hx(\tau)}p(x,\tau|x_{0},0)\int_{0}^{L}e^{-hx_{0}}dx_{0}/L\\
\equiv & \frac{1}{L}\int_{0}^{L}\langle e^{hx(\tau)}\rangle_{x_{0}}e^{-hx_{0}}dx_{0}.
\end{align}
From the main text we get 
\begin{equation}
\langle e^{hx(\tau)}\rangle_{x_{0}}=e^{\mathcal{L}\tau}e^{hx_{0}}\phi(0)=e^{hx_{0}}\left[e^{\mathcal{L}\tau}\phi(0)\right]=e^{hx_{0}}\langle e^{hx(\tau)}\rangle_{x_{0}=0},
\end{equation}
so that 
\begin{align}
\langle e^{hx_{\tau}}\rangle= & \frac{1}{L}\int_{0}^{L}\langle e^{hx(\tau)}\rangle_{x_{0}}e^{-hx_{0}}dx_{0}\nonumber \\
= & \langle e^{hx(\tau)}\rangle_{x_{0}=0}\left(\frac{1}{L}\int_{0}^{L}dx_{0}\right)\nonumber \\
= & \langle e^{hx(\tau)}\rangle_{x_{0}=0},
\end{align}
which is just the equality (\ref{Equality}) of the main text. Then
using Eq.(\ref{formal}) we are able to figure out $K_{x_{\tau}}(h)$.
The matrix $\mathcal{L}$ can always be diagonalized as $\mathcal{L}=X\Lambda X^{-1},$where
$\Lambda=\text{diag}(\lambda_{1},\lambda_{2})$ and $X$ is composed
of its eigenvectors, since it has two different eigenvalues 
\begin{align*}
\lambda_{1} & =\frac{a+b-\sqrt{(a-b)^{2}+4\gamma_{l}\gamma_{r}}}{2}\\
\lambda_{2} & =\frac{a+b+\sqrt{(a-b)^{2}+4\gamma_{l}\gamma_{r}}}{2},
\end{align*}
where $a\equiv Dh^{2}+vh-\gamma_{r}$ and $b\equiv Dh^{2}-vh-\gamma_{l}$.
As a result, the exponential of matrix $\mathcal{L}$ can be computed
using 
\begin{align}
e^{\mathcal{L}\tau}= & Xe^{\Lambda\tau}X^{-1}\nonumber \\
= & X\left(\begin{array}{cc}
e^{\lambda_{1}\tau} & 0\\
0 & e^{\lambda_{2}\tau}
\end{array}\right)X^{-1},
\end{align}
leading to the final expression of $K_{x_{\tau}}(h)$: 
\begin{equation}
K_{x_{\tau}}(h)=\frac{a+b}{2}\tau+\ln\left[\cosh\left(\frac{f}{2}\tau\right)+C\sinh\left(\frac{f}{2}\tau\right)\right],\label{Cumulant}
\end{equation}
with 
\[
f\equiv\frac{\sqrt{(a-b)^{2}+4\gamma_{l}\gamma_{r}}}{2}
\]
and 
\[
C\equiv\frac{(a-b)(\gamma_{l}-\gamma_{r})+4\gamma_{l}\gamma_{r}}{f(\gamma_{l}+\gamma_{r})}.
\]
From the above expression, it's clear that the CGF is an increasing
function of $|h|$. We have checked the validity of Eq.(\ref{Cumulant})
by generating the first and second cumulant of $x_{\tau}$ utilizing
\begin{equation}
\frac{\partial K_{x_{\tau}}(h)}{\partial h}\lvert_{h=0}=\langle x_{\tau}\rangle
\end{equation}
and 
\begin{equation}
\frac{\partial^{2}K_{x_{\tau}}(h)}{\partial h^{2}}\lvert_{h=0}=\text{Var}(x_{\tau}),
\end{equation}
which equal to their true forms (\ref{meanvalue}) and (\ref{effDiff})
in the main text. 

In what follow we calculate $\lim_{h\rightarrow\infty}F(h).$ From
the definition of $f$ and $C$, it is clear that in the large $h$
limit 
\begin{align*}
f & \sim h,\ C\sim\mathcal{O}(1)\\
\Rightarrow & \ln\left[\cosh\left(\frac{f}{2}\tau\right)+C\sinh\left(\frac{f}{2}\tau\right)\right]\sim h
\end{align*}
so that from the expression of $K_{x_{\tau}}(h)$ one can obtain
\begin{equation}
K_{x_{\tau}}(h)=D\tau h^{2}+O(h)+\text{constant},
\end{equation}
with $O(h)$ being some function of the order of $h$. Then one can
readily compute $\lim_{\lvert h\rvert\rightarrow\infty}F(h)$ as
\begin{align}
\lim_{\lvert h\rvert\rightarrow\infty}F(h)= & \lim_{\lvert h\rvert\rightarrow\infty}\frac{h^{2}}{K_{x_{\tau}}(h)-h\langle x_{\tau}\rangle}\nonumber \\
= & \lim_{\lvert h\rvert\rightarrow\infty}\frac{1}{K_{x_{\tau}}(h)/h^{2}-\langle x_{\tau}\rangle/h}\nonumber \\
= & \lim_{\lvert h\rvert\rightarrow\infty}\frac{1}{D\tau+\mathcal{O}(\lvert h\rvert^{-1})}\nonumber \\
= & \frac{1}{D\tau}.
\end{align}

\section{Numerical evidence of $\text{sup}_{h}F(h)=1/(D\tau)$}

The transport efficiency for the entropy production estimator $\frac{h^{2}\langle x_{\tau}\rangle^{2}}{\ln\langle e^{hx_{\tau}}\rangle-h\langle x_{\tau}\rangle}$
is given by 
\begin{align}
\eta(\tau,h)= & \frac{\langle x_{\tau}\rangle^{2}}{\langle\Sigma_{\tau}\rangle}F(h)\nonumber \\
= & D\tau\left(\frac{\gamma_{l}-\gamma_{r}}{\gamma_{l}+\gamma_{r}}\right)^{2}F(h),
\end{align}
thus $F(h)\propto\eta(\tau,h)$. As a result, to test the monotonicity
of $F(h)$ one could focus on $\eta(\tau,h)$. In what follow we give
the three-dimensional plots of $\eta(\tau,h)$ versus $\tau$ and
$h$ with different $v$, $\gamma_{l,r}$ and $D$, where the vertical
axes denotes $\eta(\tau,h)$. 

\begin{figure}
\begin{centering}
\includegraphics[width=1\columnwidth]{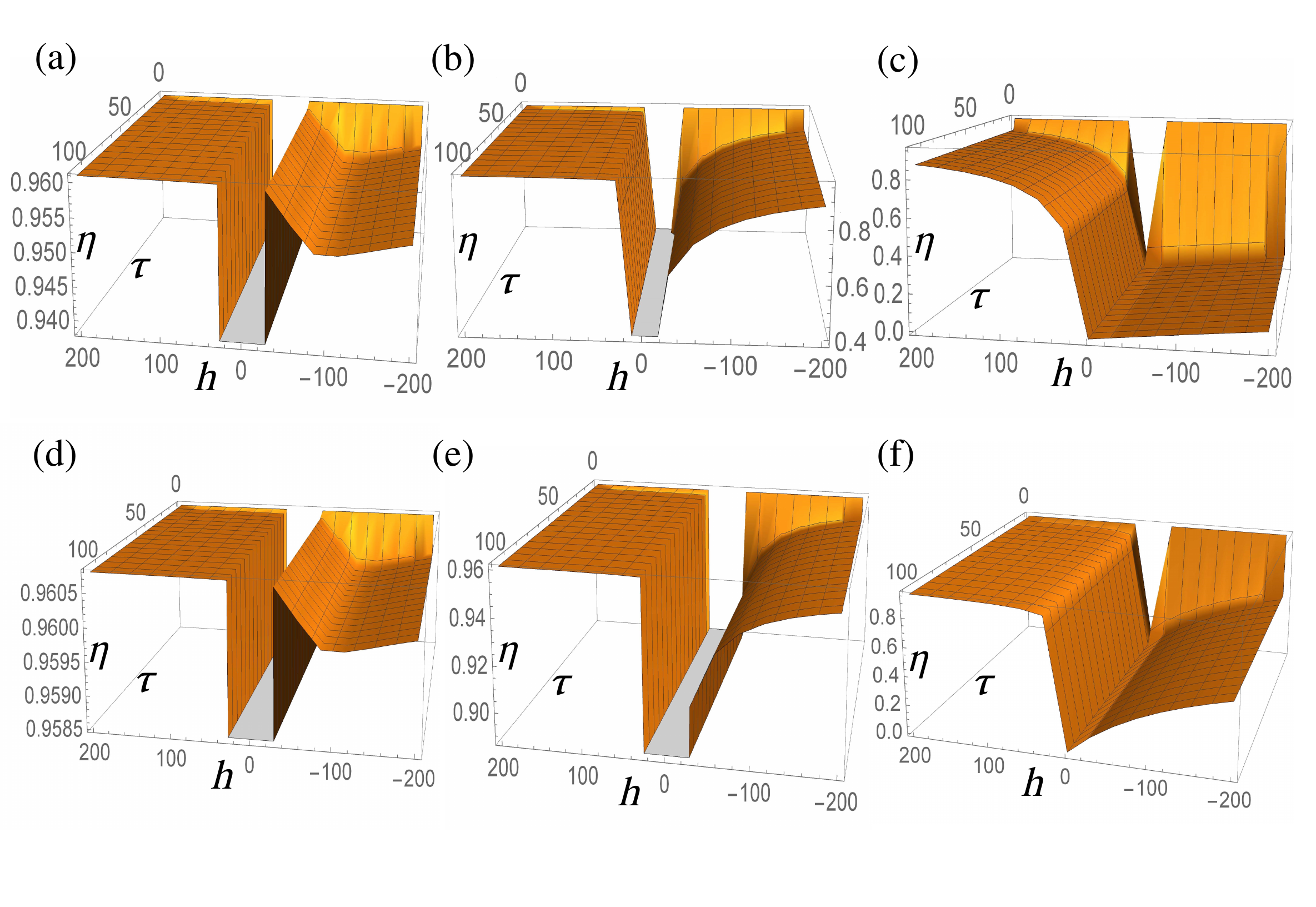}
\par\end{centering}
\caption{The transport efficiency $\eta(\tau,h)$ versus $h$ and $\tau$ with
different $D$ and $v$, transition rates are $\gamma_{l}=10,\ \gamma_{r}=0.1$
for this figure. For (a), (b) and (c) the parameters are chosen as
$D=1.0,\ v=0.1,\ 1.0,\ 100$, respectively. For (d), (e) and (f) the
parameters are chosen as $D=0.1,\ v=0.1,\ 1.0,\ 100,$ respectively. }

\label{figS1}
\end{figure}
Above plots (Fig. 4) show the behaviors of $\eta(\tau,h)$
versus $h$ and $\tau$ when $\gamma_{l}=10$ and $\gamma_{r}=0.1$.
Below we explore another case when $\gamma_{l}=5$ and $\gamma_{r}=1$.
\textcolor{red}{Note that in the gray areas of the plots, the value
of $\eta(\tau,h)$ is very close to zero compared to the values of
points in other areas.}

\begin{figure}
\begin{centering}
\includegraphics[width=1\columnwidth]{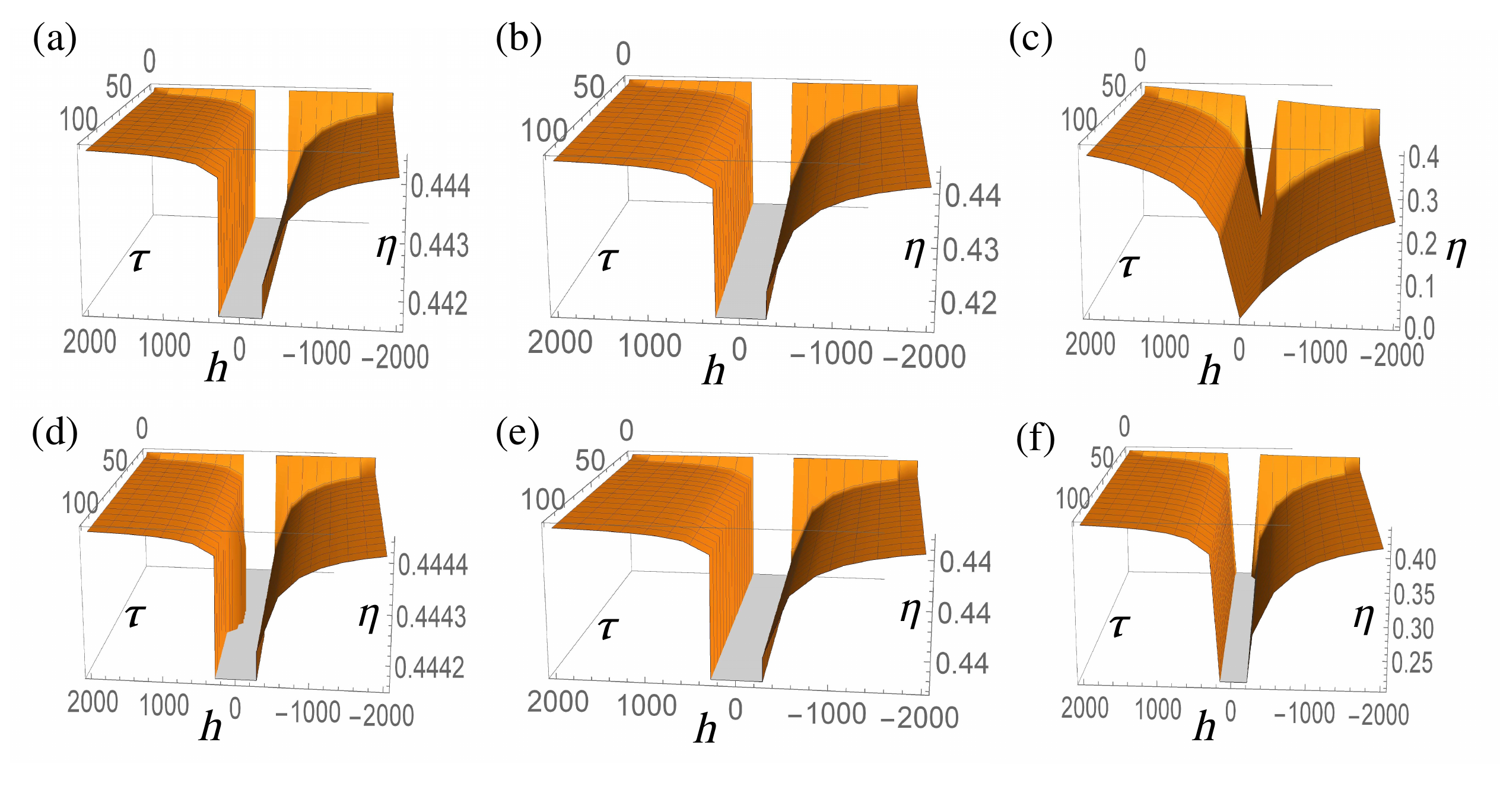}
\par\end{centering}
\caption{The transport efficiency $\eta(\tau,h)$ versus $h$ and $\tau$ with
different $D$ and $v$, transition rates are $\gamma_{l}=5,\ \gamma_{r}=1$
for this figure. For (a), (b) and (c) the parameters are chosen as
$D=1.0,\ v=0.1,\ 1.0,\ 100$, respectively. For (d), (e) and (f) the
parameters are chosen as $D=0.1,\ v=0.1,\ 1.0,\ 100,$ respectively. }

\label{figS2}
\end{figure}
With these numerical evidence, we could claim that $F(h)$ is an increasing
function of $h$ when $h>0$ and $F(h)\leq F(-h)$ when $h<0$ for
any observation time $\tau$, leading to the wanted result 
\begin{equation}
\sup_{h}F(h)=\lim_{h\rightarrow\infty}F(h)=\frac{1}{D\tau}.
\end{equation}
Note that when $\gamma_{l}=\gamma_{r},$ $F(h)$ becomes a even function
and $F(h)=F(-h)$, when $\gamma_{l}>\gamma_{r}$ we observe that $F(h)>F(-h)$. 

We also check other cases when these parameters take other values,
and no counterexample has been found.

\end{document}